\documentclass{emulateapj}
\usepackage{epsfig}		
\usepackage{amsmath}
\usepackage{subfigure}
\usepackage{color}

\shorttitle{Large-Scale Environmental Effects}
\shortauthors{Jung et al.}

\begin{document}
\slugcomment{Accepted for publication in ApJ, August 14, 2014}

\title{Effects of large-scale environment on the assembly history of central galaxies}
\author{Intae Jung$^{1,}$\altaffilmark{2}, Jaehyun Lee$^1$, and Sukyoung K. Yi$^1$}
\affil{$^1$Department of Astronomy and Yonsei University Observatory, Yonsei University, Seoul 120-749, Republic of Korea; yi@yonsei.ac.kr}
\altaffiltext{2}{Current address: Department of Astronomy, The University of Texas at Austin, Austin, TX 78712, USA}

\begin{abstract}
We examine whether large-scale environment affects the mass assembly history of their central galaxies.
To facilitate this, we constructed dark matter halo merger trees from a cosmological N-body simulation and calculated the formation and evolution of galaxies using a semi-analytic method.
We confirm earlier results that smaller halos show a notable difference in formation time with a mild dependence on large-scale environment.
However, using a semi-analytic model, we found that on average the growth rate of the stellar mass of central galaxies is largely insensitive to large-scale environment.
Although our results show that the star formation rate (SFR) and the stellar mass of central galaxies in smaller halos are slightly affected by the assembly bias of halos, those galaxies are faint, and the difference in the SFR is minute, and therefore it is challenging to detect it in real galaxies given the current observational accuracy.
Future galaxy surveys, such as the BigBOSS experiment and the Large Synoptic Survey Telescope, which are expected to provide observational data for fainter objects, will provide a chance to test our model predictions.
\end{abstract}

\keywords{methods : N-body simulations --- cosmology: large-scale structure of universe --- galaxies: halos --- galaxies: evolution}

\section{Introduction} 
In the standard $\Lambda$CDM cosmology with the hierarchical structure formation paradigm, dark matter halos not only play an important role in galaxy formation and evolution, but also compose large-scale structures in the universe.
Dark matter first collapses and forms a gravitationally bound system, then baryons fall into the pre-existing potential well of dark matter halo and form galaxies \citep{White78}. 
Small halos grow into larger systems with mergers or accretion, and thus galaxies residing in the halos naturally evolve with the halo growth.
Furthermore, in the past decade, large galaxy survey projects such as the 2dF Galaxy Redshift Survey \citep[2dFGRS;][]{Colless01} and the Sloan Digital Sky Survey \citep[SDSS;][]{York00} and large volume simulations using supercomputers like the Millennium simulation \citep{Springel05nat} and the Horizon runs \citep{kim09, kim11} have revealed the existence of large-scale structures.

Since \cite{oemler74} and \cite{dressler80} found that the cluster environment affects the galaxy color, there have been numerous studies verifying the relations between galaxy properties and their environments. 
However, most of them have investigated local environments, which can be classified into cluster, group, or field. 
While the local environments of galaxies are confined to the inner halo regions, large-scale structures, such as superclusters, filaments, and voids are super-halo scale environments.
Therefore, the large-scale environment should be distinguished from the local environments.

\cite{Mo96} predicted that massive halos are more likely to occupy denser regions in a large-scale distribution.
Moreover, \cite{Mo04} depicted that the large-scale environmental dependence of the galaxy luminosity function can be explained by the mass dependence of the halo bias. 
Galaxy properties mainly depend on the inner halo environment defined by halo mass, and the bias of halo mass in large-scale distribution can cause large-scale environmental dependence.
However, using N-body simulations, \cite{Sheth04} showed that halos in denser regions form slightly earlier than those in less dense regions, when halo mass is fixed. 
In addition, many theoretical works have found that the formation time of small halos ($\lesssim10^{13}h^{-1}M_{\odot}$ at $z=0$) depends on the large-scale environment, which is known as \textit{the assembly bias} \citep[e.g.][]{Gao05, Harker06, Wechsler06, Maulbetsch07, Jing07, Gao07, Li08, Fakhouri09, Fakhouri10a}. 
In short, not only is the halo mass biased by the large-scale environment, but the assembly time of halos also depends on the large-scale environment.

The large-scale environmental effects on galaxies have become an active research area with the advent of wide field surveys. 
\cite{Kauffmann04} and \cite{Blanton07} attempted to find a correlation between the large-scale environment and the central galaxy color in SDSS galaxy groups; however, they concluded that the large-scale environment does not have any considerable effects on the central galaxy color. 
Meanwhile, \cite{Yang06} found, from 2dFGRS data, that galaxy groups with passive central galaxies at a fixed halo mass are strongly clustered, and it was supported by \cite{Wang08} using SDSS galaxies. 
Furthermore, \cite{Croton07} studied the correlation between the halo assembly bias and galaxy clustering using a semi-analytic model \citep{Croton06} and reached a conclusion similar to the observational study of \cite{Yang06}.
\cite{scudder12} also pointed out that the star formation rate of compact group galaxies is mildly dependent on the large-scale environment, based on SDSS data.
Currently, it seems that galaxy properties are correlated with the large-scale environment in addition to the mass of host halo.
However, there is still no clear understanding of the physical origins of these large-scale environmental effects in the baryonic universe.

The aim of this paper is to study the effect of large-scale dependence of halo assembly time on the mass assembly of galaxies using a cosmological N-body simulation and a semi-analytic model. 
In addition to the previous study of \cite{Croton07}, which investigates the effects of the halo assembly bias on galaxy clustering, we examine the effect of the halo assembly bias on the mass assembly of galaxies.
We first measure the mass assembly of halos in different large-scale environments from halo merger trees, and trace the evolution of the central galaxies in those halos by using a semi-analytic model. 
This scheme allows us to study the direct connection between halos and their central galaxies according to the mass assembly history and to test the existence of any effects of the halo assembly bias on those galaxies.

This paper is organized as follows. 
In section 2, we describe the details of our cosmological N-body simulation and the scheme by which we construct halo merger trees from N-body simulations. 
In section 3, we explain the measurement of the large-scale environment of halos and the classification of our sample halos and galaxies. 
We study the mass assembly history of dark matter halos and their central galaxies in section 4, and in the final section, we discuss and summarize our results.

\section{Methodology}
\subsection{N-body simulation}
We performed a cosmological dark matter-only simulation using parameters of the standard concordance cold dark matter cosmology using a cosmological constant $\Omega_{\Lambda} = 0.728$, a current matter density parameter $\Omega_{m} = 0.272$, the normalization factor of the initial power spectrum $\sigma_{8} = 0.809$, the spectral index $n = 0.963$, and the Hubble constant $H_{0}=100h$km$s^{-1}$ with $h = 0.704$. 
These parameter values are based on the seven-year Wlikinson Microwave Anisotropy Probe observations \citep[WMAP;][]{Jarosik11}. 
The cosmological hydrodynamic code used in this work is GADGET-2 \citep{Springel05}. 
For the simulation, a periodic box-size was chosen as $100h^{-1}$Mpc with $1024^{3}$ dark matter particles.\footnotemark[1]
The dark matter particle mass was $6.9 \times 10^{8}h^{-1}M_{\odot}$, and the gravitational softening length was $2.441 h^{-1}$kpc. 
The dark matter particle position and velocity information were stored for 117 time-steps between z=13 and z=0. 
The initial condition of the simulation was produced using MPgrafics \citep{Prunet08}, which is a parallel version of Grafic1 \citep{Bertschinger01}.
\footnotetext[1]{We note that our analysis yields consistent results as shown in this paper even when we used a smaller ($70h^{-1}Mpc$) volume simulation with $512^3$ particles.}

\subsection{Dark matter halo merger trees : ySAMtm}
In order to trace the formation and the mass assembly history of halos, we built dark matter halo merger trees from an N-body simulation.
Our treebuilder, ySAMtm, was implemented into the semi-analytic model \cite[ySAM;][]{Lee13}. 

The first step in building dark matter halo merger trees is structure detection. 
We used the structure finding code AHF \citep{Knollmann09} in this work.
We define the halo center as the position of the densest region in the halo, as the center of mass is not a proper position for halos undergoing mergers. 
The halos and subhalos in the simulation are supposed to have at least 20 particles.

From the results of structure detection, we constructed dark matter halo merger trees. 
Figure \ref{fig:halo merger} briefly describes the definition of dark matter halo mergers. 
We define the beginning of the halo mergers using the virial radius criterion. 
A detailed procedure for building halo merger trees can be explained as follows :

\begin{figure}
\centering
\includegraphics[width=0.5\textwidth]{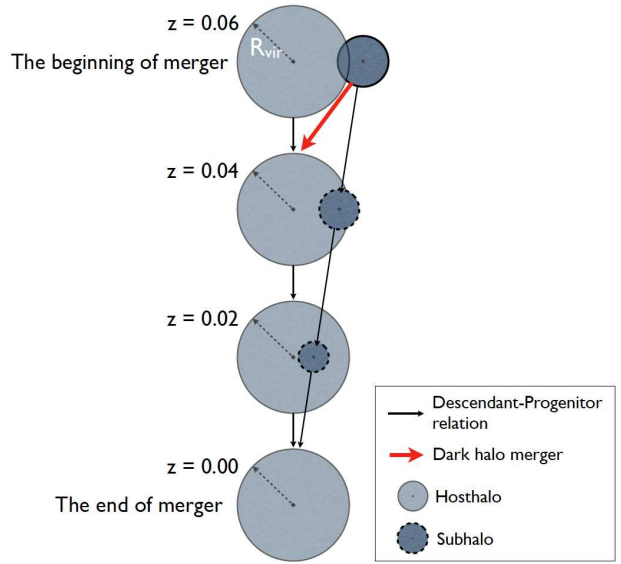}
\caption[The definition of a dark matter halo merger]{Definition of a dark matter halo merger. This schematic diagram shows the processes of a dark matter halo merger. The solid circle indicates a main halo, and the dashed circle is a subhalo. The black (thin) solid arrow describes the unique descendant-progenitor relation, and the red (thick) solid arrow states the beginning of the dark matter halo merger.\label{fig:halo merger}}
\end{figure}

\begin{enumerate}
\item \text{Calculate the shared mass between halos}

The shared mass between halos at two subsequent snapshots is calculated by tracing the particle transfer between those halos. 
At this step, particle transfers between halos in two snapshots are tracked using particle indexes, and individual particles should be included in a single halo or subhalo; particles in a subhalo are not listed in the main halo of the subhalo.
\item \text{Define a unique descendant halo}.

In order to be applicable to semi-analytic models, we find the unique descendant halos of individual progenitor halos.
For one progenitor halo, the shared masses of its descendant halos are compared, and the descendant halo with the largest shared mass among those descendants is assigned as a unique descendant halo of the progenitor. 
However, if another descendant halo receives the largest fraction of its mass from the same progenitor, we determine this descendant halo to be a unique descendant of the progenitor, even if the shared mass of the descendant is not the largest compared to the other descendants.
This additional process is needed to avoid a situation in which the small descendant halo is considered to be otherwise newly born.
Thus far, one descendant halo can be connected to multiple progenitor halos, whereas a single progenitor must have a unique descendant. 
If one descendant halo has multiple progenitors, the main progenitor halo is determined by maximizing 
$M_{shared,i}/M_{descendant}$
, where $M_{shared,i}$ is the shared mass between the descendant halo and the progenitor halo, and $i$ is the list index of the progenitor halo.
In Figure \ref{fig:halo merger}, the black (thin) solid arrows indicate a unique descendant-progenitor relation.

\item \text{Find the end of the merger events.} 

The end of the halo mergers is defined as the moment when two progenitor halos have an identical descendant halo at the next snapshot.
In Figure \ref{fig:halo merger}, the main halo and its subhalo at z=0.02 have the same descendant halo at z=0. 
In this case, we consider z=0 to be the epoch when a halo merger is complete.
\item \text{Detect the beginnings of halo mergers.} 

In order to do this, the progenitors of both the primary main halo and its subhalo are tracked backward, starting from the end of merger. 
At every output, the distance between the two progenitors is compared to the virial radius of the primary halo. 
The moment when the small halo crosses the virial radius of the main halo is considered to be the beginning of the halo merger. 
This is equivalent to the moment when the small halo merges with the larger halo and becomes its substructure. 
As seen in Figure \ref{fig:halo merger}, the small halo becomes a subhalo between z=0.06 and z=0.04, which defines the beginning of the halo merger.
\item\text{Build the merger trees of individual halos.} 

For each halo at z=0, the whole merger history is constructed by backwards tracking progenitor halos including all progenitors of the halos which have finally merged into the object halo. 
\end{enumerate}

Fly-by events between halos can significantly inflate the number of halo mergers by double counting \citep{Fakhouri08, Genel09}.
By considering only the subhalos that were supposed to be finally merged and by tracing backward, we avoided the overestimation of halo mergers caused by fly-bys.

\section{Halo Sampling}
\subsection{Measuring the environment: local overdensity}
The goal of this study is to investigate the environmental effects in large scales; therefore we need to define the large-scale environment of halos.
In the literature, there are two ways to measure the large-scale environment. 
The first is to measure the local overdensity in a sphere of radius R centered at a halo \citep[e.g.][]{Lemson99, Wang07, Maulbetsch07, Fakhouri09, Hahn09, Tillson11}, and the other is to measure the halo bias using the two-point correlation function \citep[e.g.][]{Sheth04, Gao05, Harker06, Wechsler06, Gao07, Croton07,wang13}. 
Although the two-point correlation function is widely used to study the relations between halo clustering and halo properties (formation, merger, dynamical structure, etc.), the correlation function itself is only dependent on the number of halo pairs.
On the other hand, the local overdensity more directly reflects the matter distribution of both the internal and external regions of halos.
Therefore, we chose to use the former method.

The definition of the local overdensity ($\delta_{R}$) is the same as that in \cite{Fakhouri09, Fakhouri10a} and can be calculated as follows:
\begin{eqnarray}
\delta_{R} \equiv \frac{\rho_{R} - \Bar{\rho_{m} } }{\Bar{\rho_{m} } } ,
\label{eqn:local overdensity}
\end{eqnarray}
where $\rho_{R}$ is the matter density in a sphere with radius $R$, and $\Bar{\rho_{m}}$ is the mean matter density in a simulation box. 
The contribution of the central halo mass is removed by
\begin{eqnarray}
\delta_{R-Halo} \equiv \delta_{R} - \frac{M_{Halo}}{V_{R} \Bar{\rho_{m}}},
\label{eqn:local overdensity}
\end{eqnarray}
where $M_{Halo}$ is the mass of the central halo, and $V_{R}$ is the volume of the sphere.

It is worth mentioning how we choose the sphere radius for measuring the local overdensity. 
In order to understand the impact of different sphere radii, we calculated the local overdensity with various values ($3h^{-1}$Mpc, $5h^{-1}$Mpc, $7h^{-1}$Mpc, and $9h^{-1}$Mpc). The results did not differ significantly with different sphere radii, which is consistent with previous studies \citep{Sheth04, Croton07, Maulbetsch07}.
However, similar to the results presented in \cite{Fakhouri09}, we also found that the centric distance of the farthest particle sometimes exceeded $5h^{-1}$Mpc for massive halos.
Therefore, we used $7h^{-1}$Mpc in this study.
	
\subsection{Classification}
In order to compare halos at a fixed mass in different large-scale environments, we classified the halos in the present universe into three mass bins: $10^{11.0-11.5}h^{-1}M_{\odot}$, $10^{11.5-12.0}h^{-1}M_{\odot}$, and $10^{12.0-13.0}h^{-1}M_{\odot}$.
This range of halo mass roughly covers single galaxies to small clusters. 
We focused on halos with the mass range $10^{11-13}h^{-1}M_{\odot}$ at $z=0$ because there were not enough massive halos ($\gtrsim 10^{13}h^{-1}M_{\odot}$) to be statistically analyzed, and smaller halos ($\lesssim10^{11}h^{-1}M_{\odot}$) were excluded due to the resolution limits of our simulation.
The definitions of high density and low density in large-scale environments were selected to be the top $20\%$ and the bottom $20\%$ of the local overdensity distribution ($\delta_{7-Halo}$).

The spatial distribution of halos is only slightly dependent on halo mass; the average mass of halos in high-density regions is slightly larger than that of the halos in low-density regions even at the same mass bin. 
For all identified halos with the mass range $10^{12.0-13.0}h^{-1}M_{\odot}$, the average mass of halos in denser regions (the top 20\%) is $2.73\times10^{12}h^{-1}M_{\odot}$, while that of halos in less dense regions (the bottom 20\%) is $1.87\times10^{12}h^{-1}M_{\odot}$.
To minimize the effect of halo mass dependence, we selected the same number of halos for different density groups until the average masses of halos in those groups became virtually identical.

\begin{table}
\centering
\begin{center}
\caption[Halo classification with local overdensity]{Properties of sample halos residing in different environments.\label{tab:halo classification}}
\begin{tabular}{cccc}
\tableline \tableline
\quad  	{Mass range}		&{}					      &	$M_{avg}$ in high $\delta$ 	& $M_{avg}$ in low $\delta$\\
\quad {$(h^{-1}M_{\odot})$} 	&$N_{halos}$\footnotemark[1]&	$(h^{-1}M_{\odot})$    	& $(h^{-1}M_{\odot})$	 \\
\tableline
\quad $10^{11.0-11.5}$	&	3000	& 	$1.69\times10^{11}$ 	&  	$1.69\times10^{11}$\\
\quad $10^{11.5-12.0}$	&	900   	& 	$5.34\times10^{11}$ 	&  	$5.30\times10^{11}$\\
\quad $10^{12.0-13.0}$	&	230   	& 	$1.66\times10^{12}$ 	&  	$1.59\times10^{12}$\\
\tableline
\end{tabular}
\footnotetext[1]{At a fixed mass range, the numbers of sample halos in high- and low-density groups are the same to each other.}
\tablecomments{Dark matter halos at a fixed mass range were classified into two groups according to local overdensity($\delta$). High-density and low-density environments were defined as the top 20\% and the bottom 20\% of the local overdensity distribution, respectively }
\end{center}
\end{table}

\begin{figure}
\centering
\includegraphics[width=0.5\textwidth]{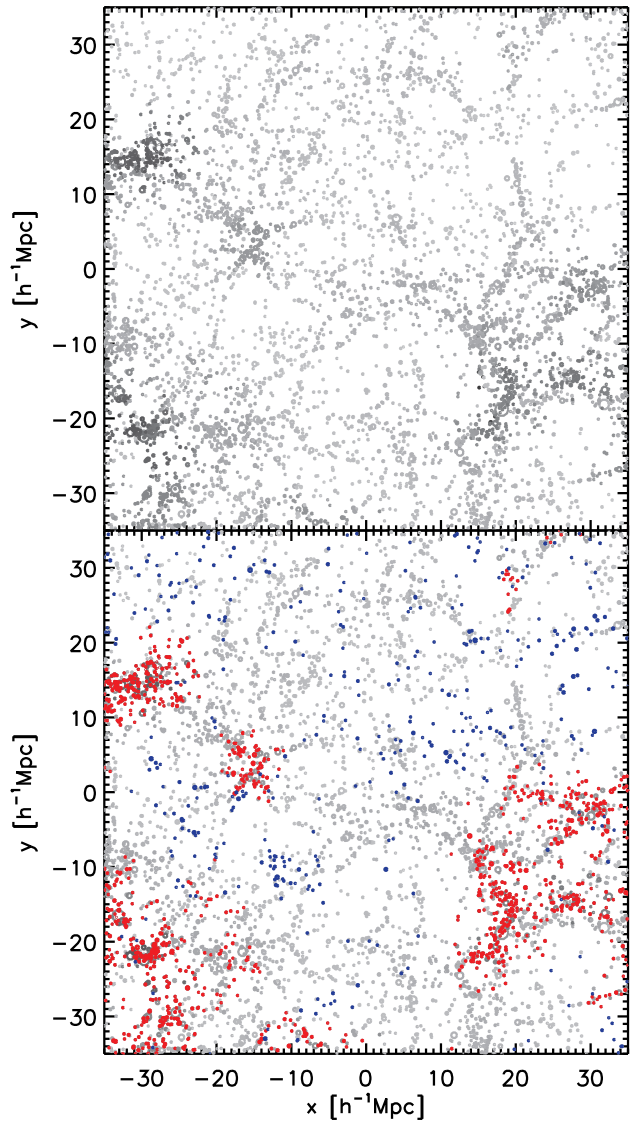}
\caption[The distribution of dark matter halos]{Dark matter halo distributions in a $30h^{-1}$Mpc slice in the simulation. The upper panel displays the distribution of all halos more massive than $10^{11}h^{-1}M_{\odot}$ in the region. The size of the circles represents the virial radius. The grayscale system represents the local overdensity $\delta_{7-Halo}$. The darker circles indicates halos in denser regions. The bottom panel is the over-plotted image of the upper panel with our sample halos in Table \ref{tab:halo classification}. The red and blue circles are halos in high-density (the upper 20\%) and low-density (the bottom 20\%) regions, respectively.
\label{fig:halo distribution}}
\end{figure}

Table \ref{tab:halo classification} presents the number and the average mass of the sample halos for each group, and Figure \ref{fig:halo distribution} illustrates the distribution of dark matter halos. As seen In Figure \ref{fig:halo distribution}, the upper panel displays the distribution of all halos more massive than $10^{11}h^{-1}M_{\odot}$ in a $30h^{-1}$Mpc slice in the simulation. 
The grayscale system represents the local overdensity $\delta_{7-Halo}$; the darker circles indicate halos in denser regions.
In the bottom panel, our sample halos are over-plotted on the upper-panel image (Table \ref{tab:halo classification}).
The red and blue circles are halos in high-density (the upper 20\%) and low-density (the bottom 20\%) regions, respectively.

The main purpose of this paper is to investigate the effects of the halo assembly bias on real galaxies.
For the purpose, we used a semi-analytic model \cite[ySAM;][]{Lee13}.
In the model, the mass accretion and mergers of galaxies follow the evolution of each halo merger tree, and many processes of baryonic physics (gas cooling, star formation, feedback processes, etc.) are calculated using simple prescriptions. 
\cite{Lee13} demonstrated that their model is calibrated to reproduce recent empirical data, and \cite{Yi13} showed that the merger history of galaxies predicted by the model is comparable to the observational results of the cluster galaxies in \cite{sheen12}.

To investigate the impact of various SAM parameters on our results presented in the parer, we varied the efficiencies of AGN and supernovae feedbacks, which are considered important mechanisms in galaxy growth, in a factor of two ranges: from half to twice of our fiducial values. These wide ranges of efficiencies made noticeable differences in the galaxy stellar mass function at z=0. We confirmed, however, that our result is not sensitive at all to the variation of feedback efficiencies.

We perform our test on {\em central} galaxies because our research focus is on the assembly bias of host halos reflecting the outer environments of those halos, and central galaxies directly follow the formation and the mass growth of their host halos.
Satellite galaxies, on the other hand, are more likely to be affected by their host halo environment.

\section{Mass assembly history}
\subsection{Dark matter halos and gaseous components}
In the current galaxy formation theory, the gas accretion in galaxies is strongly affected by the assembly history of dark matter halos. In this section, we present the assembly history of halos and their central galaxies, comparing the average values between different density environments at a fixed halo mass, as shown in Table \ref{tab:halo classification}.

\begin{figure}
\centering
\includegraphics[width=0.45\textwidth]{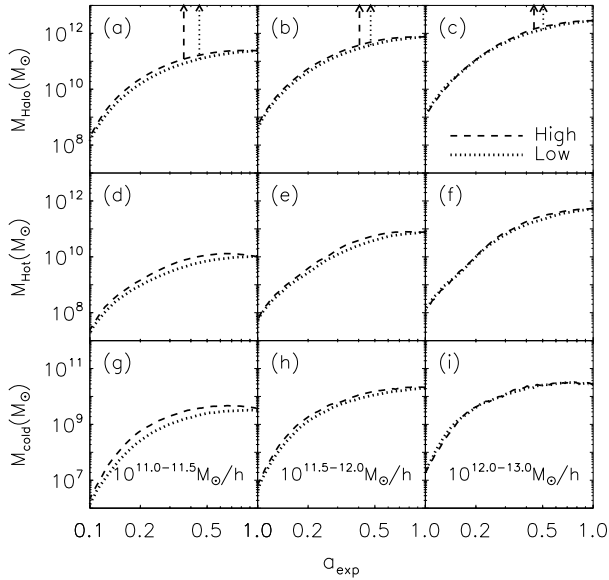}
\caption[the growth of dark matter halos and galaxy gaseous components]{The mass growth of halos and galaxy gaseous components. The average masses of dark matter halo (top row), hot gas (middle), and cold gas (bottom) components were calculated according to the different halo mass bins shown in Table \ref{tab:halo classification}: $10^{11.0-11.5}h^{-1}M_{\odot}$, $10^{11.5-12.0}h^{-1}M_{\odot}$, and $10^{12.0-13.0}h^{-1}M_{\odot}$ (from left to right). The dashed and dotted curves indicate halos in different density environments: high-density (the top 20\%) and low-density (the bottom 20\%) regions. The vertical arrows present the time at which the mass reaches half of the final mass for each component. The x-axis indicates the cosmic expansion factor. \label{fig:fig3}}
\end{figure}

Figure \ref{fig:fig3} shows the average mass growth of dark matter halo (top row), hot gas (middle), and cold gas (bottom) components. 
The average masses are compared in three different halo mass bins: $10^{11.0-11.5}h^{-1}M_{\odot}$, $10^{11.5-12.0}h^{-1}M_{\odot}$, and $10^{12.0-13.0}h^{-1}M_{\odot}$ at $z=0$ (from left to right). 
The dashed curve was calculated using halos in a high-density environment (the top 20\%), and the dotted curve was computed using those in a low-density environment (the bottom 20\%). 
The vertical arrows indicate the formation time\footnote{Conventionally, the halo formation time is defined as the time at which the mass of the main progenitor halo becomes larger than half of the final descendant halo mass. We followed this definition.} of dark matter halos.

In the top row of Figure \ref{fig:fig3}, the average mass of the halos shows that halos form earlier in denser regions than in less dense regions, and the difference in the halo formation time is larger in lower-mass halos (panel (a)).
This is in agreements with the previous studies \citep[e.g.][]{Gao05, Harker06, Wechsler06, Fakhouri10a}. 
In the case of gaseous components, the growth trend of hot gas (middle row) is very similar to that of dark matter halos, and the cold gas component (bottom row) follows the trend of dark matter halos and hot gas. 
In the semi-analytic model, the amount of hot gas accreted onto dark matter halos follows the global baryonic fraction, $\Omega_{b}/\Omega_{m}$, and cold gas increases from the cooling of the hot gas component.
Thus, the mass growth of hot and cold gas is similar to that of dark matter halos.

\subsection{Star formation rate}
\begin{figure}
\centering
\includegraphics[width=0.45\textwidth]{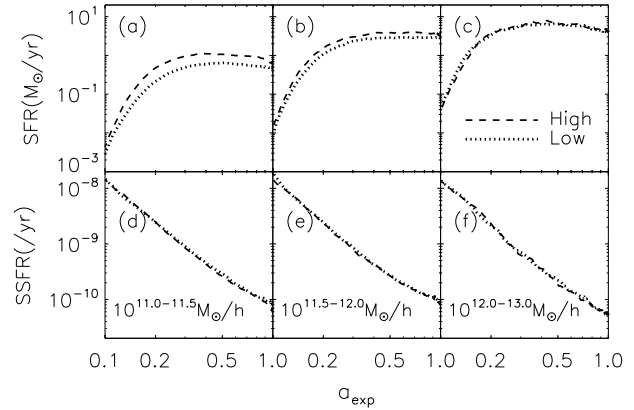}
\caption[the star formation rates and the growth of stellar mass]{The star formation rates (SFRs; top row) and the specific star formation rates (SSFRs; bottom) in the same halo mass bins as seen in Figure \ref{fig:fig3}. The dashed curve was calculated using halos in a high-density environment (the top 20\%), and the dotted curve was computed using those in a low-density environment (the bottom 20\%). The x-axis indicates the cosmic expansion factor. \label{fig:fig4}}
\end{figure}

Following the growth history of gaseous components, we investigate the star formation history and the mass growth of stellar mass in central galaxies. 
Figure \ref{fig:fig4} shows the star formation rates (SFRs; upper row) and the specific star formation rates (SSFRs; bottom) in the same mass bins as in Figure \ref{fig:fig3}. 

In the top row, the SFRs of central galaxies in smaller halos ($\lesssim 10^{12}h^{-1}M_{\odot}$) are slightly higher in denser regions than in less dense regions, although they are almost the same in the case of more massive halos.
This is because the central galaxies in smaller halos have more cold gas in high-density regions than in low-density regions (see panel (g) in Figure \ref{fig:fig3}), and the SFR is strongly dependent on the amount of cold gas. 

On the other hand, the SSFRs, the normalized SFRs, shows the same feature in both high- and low-density environments in all of the mass bins (the bottom panels in Figure \ref{fig:fig4}), which indicates that the amount of the newly formed stars per unit stellar mass is almost the same.
This is consistent with a previous observational study on the star formation history by \cite{Kauffmann04} who concluded that the SSFR minimally depends on environments larger than $1$Mpc from a galaxy.

\subsection{Mass growth rates of halos and central galaxies}
This section describes the mass growth rates of both halos and their central galaxies in an attempt to more directly search for the evidence of the assembly bias in the baryonic universe. 
\cite{Maulbetsch07} investigated halo mass growth using an N-body simulation and concluded that, at present, low-mass halos are more apt to accrete their mass in low-density regions than in high-density regions, because halos in high-density regions form earlier. 
This means that halos in denser regions (e.g., filaments) grow faster than those in less dense region (e.g., voids) when halos mass is fixed.

\begin{figure}
\centering
\includegraphics[width=0.45\textwidth]{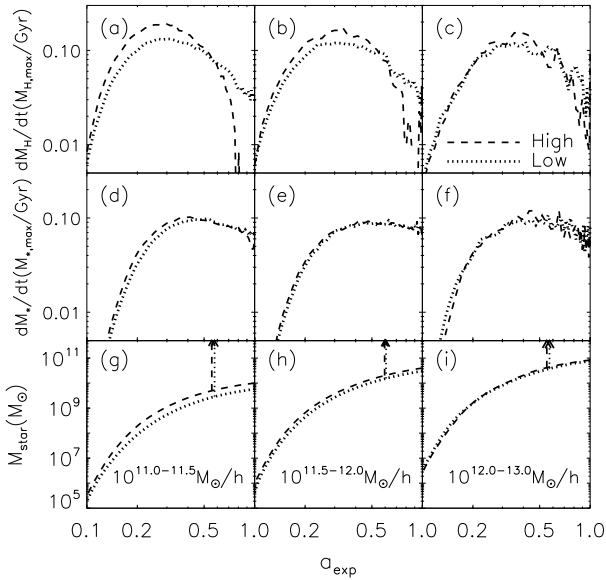}
\caption[the normalized mass growth rates of halos and central galaxies]{The normalized mass growth rates of halos (top row) and central galaxies (middle), and the stellar mass (bottom) in the same halo mass bins as shown in Figure \ref{fig:fig3}. The dashed curve was calculated using halos in a high-density environment (the top 20\%), and the dotted curve was computed using those in a low-density environment (the bottom 20\%). The vertical arrows indicate the time at which the stellar mass reaches half of the final stellar mass of central galaxies. The x-axis indicates the cosmic expansion factor. \label{fig:fig5}}
\end{figure}

As seen in the first row of Figure \ref{fig:fig5}, we reconfirmed that smaller halos in high-density regions grow faster.
However, based on the mass growth rates of central galaxies (middle row), we cannot find evidence of any considerable dependence on the large-scale environment in any mass bins. 
In particular, in low-mass halos where mass growth shows a stronger assembly bias, the difference between the mass growth rates becomes mostly diminished in galaxies (panel (d)). 

The panels in the bottom row show the stellar mass of central galaxies. 
The galaxy formation times (vertical arrows) are not sensitive to the large-scale environments, even in smaller halos, where halos show a relatively stronger bias for assembly time (see panel (a) of figure \ref{fig:fig3}). 
However, as a consequence of the higher SFR in small halos, stellar mass is slightly larger in denser regions than in less dense regions (panel (g)).

\section{Summary and Discussion}
This research attempted to address how the assembly bias by the outer environment of halos affects the mass assembly of their central galaxies. It seems that the assembly time of dark matter halos is affected by the surrounding large-scale environment in the case of small halos that mainly host isolated galaxies. 
Since the galaxy properties are strongly governed by the mass assembly of halos, it is supposed that the assembly bias probably affects the galaxies in various ways.

In short, we predict that the SSFR and the normalized growth rate of the stellar mass show no marked sign of the assembly bias for any
mass range. 
Meanwhile, the stellar masses of central galaxies in fixed mass halos are slightly affected by the assembly bias of the halos. 
However, this effect is perhaps too small to detect in real galaxies, in light of the fact that, in observational data, accurate measurement of the halo mass of galaxies is problematic \citep{Yang06}.
 
In this paper, we investigated whether central galaxies show the bias of mass assembly due to large-scale environments when considering over the effect of halo mass using a cosmological simulation and a semi-analytic model. 
Our study is based on a controlled sample of halos and galaxies, which minimizes the dependence of halo mass. 
The results can be summarized as follows:
\begin{enumerate}
\item[$\bullet$]  Using an N-body simulation, we found that the assembly time of smaller halos ($\lesssim 10^{12}h^{-1}M_{\odot}$) at a fixed mass depends on the large-scale environment. 
Halos in denser regions tend to form slightly earlier than those in sparse regions, for a given halo mass. 
\item[$\bullet$] The SSFRs and the normalized growth rates of the stellar mass of central galaxies hardly depend on the large-scale environment for all mass bins. 
For example, according to our models, the average values of the stellar 
mass-weighted 
age of central galaxies in those 
small 
halos in high- and low-density regions are $6.4\pm1.4$Gyr and $6.1\pm1.2$Gyr, respectively.
\item[$\bullet$] In the case of 
small
halos ($\lesssim10^{12}h^{-1}M_{\odot}$), the assembly bias of halos due to the large-scale environment mildly affects the SFRs and the stellar masses of central galaxies in those halos.
The SFRs of central galaxies are slightly higher in denser regions than in less dense regions, and consequently, the stellar masses of central galaxies are 
larger
in high-density regions.
\end{enumerate}

Even in the case of small halos ($10^{11.0-11.5}h^{-1}M_{\odot}$), which show a relatively stronger assembly bias, we predict that the growth rate of the stellar mass will not easily show the assembly bias. 
This is comprehensible in the sense that, unlike the assembly history of halos, galaxies are affected by complex baryonic physics such as gas cooling, feedback processes, and cluster environmental effects. 
Those baryonic physics operate differently in individual galaxies through cosmic time, which mostly erases the records of the differences in the halo formation time.

Recently, \cite{Wang13} showed that the clustering of central galaxies depends on the SSFR by using two-point correlation function statistics (see their Figure 2): low-SSFR galaxies being more strongly clustered. 
At first glance, it seemed to be in contradiction to our result, but in truth it is not. 
The discrepancy arises from the use of different approaches, and our result and theirs show the two sides of the same coin.
\cite{Wang13} picked high and low SSFR galaxies and compared their spatial correlations.
It is generally believed that at a fixed stellar mass, lower SSFR galaxies at $z=0$ tend to have earlier formation times. 
Thus, their SSFR-based classification is inherently based on \textit{galaxy assembly time}.
We on the other hand classified galaxies by \textit{large-scale density environment}.
In our approach, sample galaxies in the same density environment have a variety of star formation histories, and the galaxies showing a strong assembly bias account for a relatively small fraction. 
As a result of our approach, the mean SFRs and final stellar masses of central galaxies in the different environments show up to a factor of two difference, which is clearly present but challengingly small to be detected in currently available observational data.
Future galaxy surveys such as the BigBOSS experiment \citep{bigboss} and the Large Synoptic Survey Telescope \citep[LSST;][]{lsst}, which are expected to be able to detect fainter objects, and more accurate measurements of host halo mass will provide a better chance to test these model predictions.

\acknowledgments
SKY acknowledges support from National Research Foundation of Korea 
(Doyak FY2014). Numerical simulation 
was performed using the KISTI supercomputer under the program of KSC-2013-C3-015. 
IJ performed numerical simulations. 
JL calculated semi-analytic models. 
IJ and SKY wrote the paper.
SKY acted as the corresponding author.

\bibliographystyle{apj}

\begin{thebibliography}{44}
\expandafter\ifx\csname natexlab\endcsname\relax\def\natexlab#1{#1}\fi

\bibitem[{{Bertschinger}(2001)}]{Bertschinger01}
{Bertschinger}, E. 2001, \apjs, 137, 1

\bibitem[{{Blanton} \& {Berlind}(2007)}]{Blanton07}
{Blanton}, M.~R., \& {Berlind}, A.~A. 2007, \apj, 664, 791

\bibitem[{{Colless} {et~al.}(2001){Colless}, {Dalton}, {Maddox}, {Sutherland},
  {Norberg}, {Cole}, {Bland-Hawthorn}, {Bridges}, {Cannon}, {Collins}, {Couch},
  {Cross}, {Deeley}, {De Propris}, {Driver}, {Efstathiou}, {Ellis}, {Frenk},
  {Glazebrook}, {Jackson}, {Lahav}, {Lewis}, {Lumsden}, {Madgwick}, {Peacock},
  {Peterson}, {Price}, {Seaborne}, \& {Taylor}}]{Colless01}
{Colless}, M., {Dalton}, G., {Maddox}, S., {et~al.} 2001, \mnras, 328, 1039

\bibitem[{{Croton} {et~al.}(2007){Croton}, {Gao}, \& {White}}]{Croton07}
{Croton}, D.~J., {Gao}, L., \& {White}, S.~D.~M. 2007, \mnras, 374, 1303

\bibitem[{{Croton} {et~al.}(2006){Croton}, {Springel}, {White}, {De Lucia},
  {Frenk}, {Gao}, {Jenkins}, {Kauffmann}, {Navarro}, \& {Yoshida}}]{Croton06}
{Croton}, D.~J., {Springel}, V., {White}, S.~D.~M., {et~al.} 2006, \mnras, 365,
  11

\bibitem[{{Dressler}(1980)}]{dressler80}
{Dressler}, A. 1980, \apj, 236, 351

\bibitem[{{Fakhouri} \& {Ma}(2008)}]{Fakhouri08}
{Fakhouri}, O., \& {Ma}, C.-P. 2008, \mnras, 386, 577

\bibitem[{{Fakhouri} \& {Ma}(2009)}]{Fakhouri09}
---. 2009, \mnras, 394, 1825

\bibitem[{{Fakhouri} \& {Ma}(2010)}]{Fakhouri10a}
---. 2010, \mnras, 401, 2245

\bibitem[{{Gao} {et~al.}(2005){Gao}, {Springel}, \& {White}}]{Gao05}
{Gao}, L., {Springel}, V., \& {White}, S.~D.~M. 2005, \mnras, 363, L66

\bibitem[{{Gao} \& {White}(2007)}]{Gao07}
{Gao}, L., \& {White}, S.~D.~M. 2007, \mnras, 377, L5

\bibitem[{{Genel} {et~al.}(2009){Genel}, {Genzel}, {Bouch{\'e}}, {Naab}, \&
  {Sternberg}}]{Genel09}
{Genel}, S., {Genzel}, R., {Bouch{\'e}}, N., {Naab}, T., \& {Sternberg}, A.
  2009, \apj, 701, 2002

\bibitem[{{Hahn} {et~al.}(2009){Hahn}, {Porciani}, {Dekel}, \&
  {Carollo}}]{Hahn09}
{Hahn}, O., {Porciani}, C., {Dekel}, A., \& {Carollo}, C.~M. 2009, \mnras, 398,
  1742

\bibitem[{{Harker} {et~al.}(2006){Harker}, {Cole}, {Helly}, {Frenk}, \&
  {Jenkins}}]{Harker06}
{Harker}, G., {Cole}, S., {Helly}, J., {Frenk}, C., \& {Jenkins}, A. 2006,
  \mnras, 367, 1039

\bibitem[{{Ivezic} {et~al.}(2008){Ivezic}, {Axelrod}, {Brandt}, {Burke},
  {Claver}, {Connolly}, {Cook}, {Gee}, {Gilmore}, {Jacoby}, {Jones}, {Kahn},
  {Kantor}, {Krabbendam}, {Lupton}, {Monet}, {Pinto}, {Saha}, {Schalk},
  {Schneider}, {Strauss}, {Stubbs}, {Sweeney}, {Szalay}, {Thaler}, {Tyson}, \&
  {LSST Collaboration}}]{lsst}
{Ivezic}, Z., {Axelrod}, T., {Brandt}, W.~N., {et~al.} 2008, Serbian
  Astronomical Journal, 176, 1

\bibitem[{{Jarosik} {et~al.}(2011){Jarosik}, {Bennett}, {Dunkley}, {Gold},
  {Greason}, {Halpern}, {Hill}, {Hinshaw}, {Kogut}, {Komatsu}, {Larson},
  {Limon}, {Meyer}, {Nolta}, {Odegard}, {Page}, {Smith}, {Spergel}, {Tucker},
  {Weiland}, {Wollack}, \& {Wright}}]{Jarosik11}
{Jarosik}, N., {Bennett}, C.~L., {Dunkley}, J., {et~al.} 2011, \apjs, 192, 14

\bibitem[{{Jing} {et~al.}(2007){Jing}, {Suto}, \& {Mo}}]{Jing07}
{Jing}, Y.~P., {Suto}, Y., \& {Mo}, H.~J. 2007, \apj, 657, 664

\bibitem[{{Kauffmann} {et~al.}(2004){Kauffmann}, {White}, {Heckman},
  {M{\'e}nard}, {Brinchmann}, {Charlot}, {Tremonti}, \&
  {Brinkmann}}]{Kauffmann04}
{Kauffmann}, G., {White}, S.~D.~M., {Heckman}, T.~M., {et~al.} 2004, \mnras,
  353, 713

\bibitem[{{Kim} {et~al.}(2009){Kim}, {Park}, {Gott}, \& {Dubinski}}]{kim09}
{Kim}, J., {Park}, C., {Gott}, III, J.~R., \& {Dubinski}, J. 2009, \apj, 701,
  1547

\bibitem[{{Kim} {et~al.}(2011){Kim}, {Park}, {Rossi}, {Lee}, \& {Gott}}]{kim11}
{Kim}, J., {Park}, C., {Rossi}, G., {Lee}, S.~M., \& {Gott}, III, J.~R. 2011,
  Journal of Korean Astronomical Society, 44, 217

\bibitem[{{Knollmann} \& {Knebe}(2009)}]{Knollmann09}
{Knollmann}, S.~R., \& {Knebe}, A. 2009, \apjs, 182, 608

\bibitem[{{Lee} \& {Yi}(2013)}]{Lee13}
{Lee}, J., \& {Yi}, S.~K. 2013, \apj, 766, 38

\bibitem[{{Lemson} \& {Kauffmann}(1999)}]{Lemson99}
{Lemson}, G., \& {Kauffmann}, G. 1999, \mnras, 302, 111

\bibitem[{{Li} {et~al.}(2008){Li}, {Mo}, \& {Gao}}]{Li08}
{Li}, Y., {Mo}, H.~J., \& {Gao}, L. 2008, \mnras, 389, 1419

\bibitem[{{Maulbetsch} {et~al.}(2007){Maulbetsch}, {Avila-Reese},
  {Col{\'{\i}}n}, {Gottl{\"o}ber}, {Khalatyan}, \& {Steinmetz}}]{Maulbetsch07}
{Maulbetsch}, C., {Avila-Reese}, V., {Col{\'{\i}}n}, P., {et~al.} 2007, \apj,
  654, 53

\bibitem[{{Mo} \& {White}(1996)}]{Mo96}
{Mo}, H.~J., \& {White}, S.~D.~M. 1996, \mnras, 282, 347

\bibitem[{{Mo} {et~al.}(2004){Mo}, {Yang}, {van den Bosch}, \& {Jing}}]{Mo04}
{Mo}, H.~J., {Yang}, X., {van den Bosch}, F.~C., \& {Jing}, Y.~P. 2004, \mnras,
  349, 205

\bibitem[{{Oemler}(1974)}]{oemler74}
{Oemler}, Jr., A. 1974, \apj, 194, 1

\bibitem[{{Prunet} {et~al.}(2008){Prunet}, {Pichon}, {Aubert}, {Pogosyan},
  {Teyssier}, \& {Gottloeber}}]{Prunet08}
{Prunet}, S., {Pichon}, C., {Aubert}, D., {et~al.} 2008, \apjs, 178, 179

\bibitem[{{Schlegel} {et~al.}(2011){Schlegel}, {Abdalla}, {Abraham}, {Ahn},
  {Allende Prieto}, {Annis}, {Aubourg}, {Azzaro}, {Baltay}, {Baugh}, {Bebek},
  {Becerril}, {Blanton}, {Bolton}, {Bromley}, {Cahn}, {Carton},
  {Cervantes-Cota}, {Chu}, {Cortes}, {Dawson}, {Dey}, {Dickinson}, {Diehl},
  {Doel}, {Ealet}, {Edelstein}, {Eppelle}, {Escoffier}, {Evrard}, {Faccioli},
  {Frenk}, {Geha}, {Gerdes}, {Gondolo}, {Gonzalez-Arroyo}, {Grossan},
  {Heckman}, {Heetderks}, {Ho}, {Honscheid}, {Huterer}, {Ilbert}, {Ivans},
  {Jelinsky}, {Jing}, {Joyce}, {Kennedy}, {Kent}, {Kieda}, {Kim}, {Kim},
  {Kneib}, {Kong}, {Kosowsky}, {Krishnan}, {Lahav}, {Lampton}, {LeBohec}, {Le
  Brun}, {Levi}, {Li}, {Liang}, {Lim}, {Lin}, {Linder}, {Lorenzon}, {de la
  Macorra}, {Magneville}, {Malina}, {Marinoni}, {Martinez}, {Majewski},
  {Matheson}, {McCloskey}, {McDonald}, {McKay}, {McMahon}, {Menard},
  {Miralda-Escude}, {Modjaz}, {Montero-Dorta}, {Morales}, {Mostek}, {Newman},
  {Nichol}, {Nugent}, {Olsen}, {Padmanabhan}, {Palanque-Delabrouille}, {Park},
  {Peacock}, {Percival}, {Perlmutter}, {Peroux}, {Petitjean}, {Prada},
  {Prieto}, {Prochaska}, {Reil}, {Rockosi}, {Roe}, {Rollinde}, {Roodman},
  {Ross}, {Rudnick}, {Ruhlmann-Kleider}, {Sanchez}, {Sawyer}, {Schimd},
  {Schubnell}, {Scoccimaro}, {Seljak}, {Seo}, {Sheldon}, {Sholl},
  {Shulte-Ladbeck}, {Slosar}, {Smith}, {Smoot}, {Springer}, {Stril}, {Szalay},
  {Tao}, {Tarle}, {Taylor}, {Tilquin}, {Tinker}, {Valdes}, {Wang}, {Wang},
  {Weaver}, {Weinberg}, {White}, {Wood-Vasey}, {Yang}, {Yeche}, {Zakamska},
  {Zentner}, {Zhai}, \& {Zhang}}]{bigboss}
{Schlegel}, D., {Abdalla}, F., {Abraham}, T., {et~al.} 2011, arXiv:1106.1706

\bibitem[{{Scudder} {et~al.}(2012){Scudder}, {Ellison}, \&
  {Mendel}}]{scudder12}
{Scudder}, J.~M., {Ellison}, S.~L., \& {Mendel}, J.~T. 2012, \mnras, 423, 2690

\bibitem[{{Sheen} {et~al.}(2012){Sheen}, {Yi}, {Ree}, \& {Lee}}]{sheen12}
{Sheen}, Y.-K., {Yi}, S.~K., {Ree}, C.~H., \& {Lee}, J. 2012, \apjs, 202, 8

\bibitem[{{Sheth} \& {Tormen}(2004)}]{Sheth04}
{Sheth}, R.~K., \& {Tormen}, G. 2004, \mnras, 350, 1385

\bibitem[{{Springel}(2005)}]{Springel05}
{Springel}, V. 2005, \mnras, 364, 1105

\bibitem[{{Springel} {et~al.}(2005){Springel}, {White}, {Jenkins}, {Frenk},
  {Yoshida}, {Gao}, {Navarro}, {Thacker}, {Croton}, {Helly}, {Peacock}, {Cole},
  {Thomas}, {Couchman}, {Evrard}, {Colberg}, \& {Pearce}}]{Springel05nat}
{Springel}, V., {White}, S.~D.~M., {Jenkins}, A., {et~al.} 2005, \nat, 435, 629

\bibitem[{{Tillson} {et~al.}(2011){Tillson}, {Miller}, \&
  {Devriendt}}]{Tillson11}
{Tillson}, H., {Miller}, L., \& {Devriendt}, J. 2011, \mnras, 417, 666

\bibitem[{{Wang} {et~al.}(2007){Wang}, {Mo}, \& {Jing}}]{Wang07}
{Wang}, H.~Y., {Mo}, H.~J., \& {Jing}, Y.~P. 2007, \mnras, 375, 633

\bibitem[{{Wang} {et~al.}(2013){Wang}, {Weinmann}, {De Lucia}, \&
  {Yang}}]{Wang13}
{Wang}, L., {Weinmann}, S.~M., {De Lucia}, G., \& {Yang}, X. 2013, \mnras, 433,
  515

\bibitem[{{Wang} {et~al.}(2008){Wang}, {Yang}, {Mo}, {van den Bosch},
  {Weinmann}, \& {Chu}}]{Wang08}
{Wang}, Y., {Yang}, X., {Mo}, H.~J., {et~al.} 2008, \apj, 687, 919

\bibitem[{{Wechsler} {et~al.}(2006){Wechsler}, {Zentner}, {Bullock},
  {Kravtsov}, \& {Allgood}}]{Wechsler06}
{Wechsler}, R.~H., {Zentner}, A.~R., {Bullock}, J.~S., {Kravtsov}, A.~V., \&
  {Allgood}, B. 2006, \apj, 652, 71

\bibitem[{{White} \& {Rees}(1978)}]{White78}
{White}, S.~D.~M., \& {Rees}, M.~J. 1978, \mnras, 183, 341

\bibitem[{{Yang} {et~al.}(2006){Yang}, {Mo}, \& {van den Bosch}}]{Yang06}
{Yang}, X., {Mo}, H.~J., \& {van den Bosch}, F.~C. 2006, \apjl, 638, L55

\bibitem[{{Yi} {et~al.}(2013){Yi}, {Lee}, {Jung}, {Ji}, \& {Sheen}}]{Yi13}
{Yi}, S.~K., {Lee}, J., {Jung}, I., {Ji}, I., \& {Sheen}, Y.-K. 2013, \aap,
  554, A122

\bibitem[{{York} {et~al.}(2000){York}, {Adelman}, {Anderson}, {Anderson},
  {Annis}, {Bahcall}, {Bakken}, {Barkhouser}, {Bastian}, {Berman}, {Boroski},
  {Bracker}, {Briegel}, {Briggs}, {Brinkmann}, {Brunner}, {Burles}, {Carey},
  {Carr}, {Castander}, {Chen}, {Colestock}, {Connolly}, {Crocker}, {Csabai},
  {Czarapata}, {Davis}, {Doi}, {Dombeck}, {Eisenstein}, {Ellman}, {Elms},
  {Evans}, {Fan}, {Federwitz}, {Fiscelli}, {Friedman}, {Frieman}, {Fukugita},
  {Gillespie}, {Gunn}, {Gurbani}, {de Haas}, {Haldeman}, {Harris}, {Hayes},
  {Heckman}, {Hennessy}, {Hindsley}, {Holm}, {Holmgren}, {Huang}, {Hull},
  {Husby}, {Ichikawa}, {Ichikawa}, {Ivezi{\'c}}, {Kent}, {Kim}, {Kinney},
  {Klaene}, {Kleinman}, {Kleinman}, {Knapp}, {Korienek}, {Kron}, {Kunszt},
  {Lamb}, {Lee}, {Leger}, {Limmongkol}, {Lindenmeyer}, {Long}, {Loomis},
  {Loveday}, {Lucinio}, {Lupton}, {MacKinnon}, {Mannery}, {Mantsch}, {Margon},
  {McGehee}, {McKay}, {Meiksin}, {Merelli}, {Monet}, {Munn}, {Narayanan},
  {Nash}, {Neilsen}, {Neswold}, {Newberg}, {Nichol}, {Nicinski}, {Nonino},
  {Okada}, {Okamura}, {Ostriker}, {Owen}, {Pauls}, {Peoples}, {Peterson},
  {Petravick}, {Pier}, {Pope}, {Pordes}, {Prosapio}, {Rechenmacher}, {Quinn},
  {Richards}, {Richmond}, {Rivetta}, {Rockosi}, {Ruthmansdorfer}, {Sandford},
  {Schlegel}, {Schneider}, {Sekiguchi}, {Sergey}, {Shimasaku}, {Siegmund},
  {Smee}, {Smith}, {Snedden}, {Stone}, {Stoughton}, {Strauss}, {Stubbs},
  {SubbaRao}, {Szalay}, {Szapudi}, {Szokoly}, {Thakar}, {Tremonti}, {Tucker},
  {Uomoto}, {Vanden Berk}, {Vogeley}, {Waddell}, {Wang}, {Watanabe},
  {Weinberg}, {Yanny}, {Yasuda}, \& {SDSS Collaboration}}]{York00}
{York}, D.~G., {Adelman}, J., {Anderson}, Jr., J.~E., {et~al.} 2000, \aj, 120,
  1579

\end{thebibliography}

\end{document}